\documentclass{icrc2009}

\usepackage{graphicx}   
\usepackage{caption}    
\usepackage[font=footnotesize]{subfig} 
\usepackage{fixltx2e}
\usepackage{url}

\newcommand{\shorttitle}[1]%
{\markboth{Proceedings of the 31\MakeLowercase{$^{st}$} ICRC, {\L}\'{o}d\'{z} 2009}{#1} }
\newcommand{\etal}{\MakeLowercase{\textit{et al. }}} 

\begin{document}
\title{The possible feature of the energy spectrum of the primary cosmic rays at ultra-high energies}

\author{\IEEEauthorblockN{L.G. Dedenko\IEEEauthorrefmark{1},
			  A.V. Glushkov\IEEEauthorrefmark{2},
                          G.F. Fedorova\IEEEauthorrefmark{1},
                          S.P. Knurenko\IEEEauthorrefmark{2},
                          I.T. Makarov\IEEEauthorrefmark{2},\\
                          D.A. Podgrudkov\IEEEauthorrefmark{1},
                          M.I. Pravdin\IEEEauthorrefmark{2},
                          T.M. Roganova\IEEEauthorrefmark{1}  and
                          I.Ye. Sleptzov\IEEEauthorrefmark{2}}
                            \\
\IEEEauthorblockA{\IEEEauthorrefmark{1}D.V. Skobeltsyn Institute of Nuclear Physics, MSU,\\ Leninskie Gory,
119992 Moscow, Russia}
\IEEEauthorblockA{\IEEEauthorrefmark{2}Yu.G. Shafer Institute of Cosmophysical Research and Aeronomy, \\
31 Lenin Ave, 677891 Yakutsk, Russia}}

\shorttitle{L.G. Dedenko \etal The possible feature...}
\maketitle

\begin{abstract}

The energies of the most energetic extensive air showers observed at the Yakutsk array have been estimated with help
of the all detectors readings instead of using of the standard procedure with a parameter $s(600)$.
These detector readings have been compared with the detector responses, calculated for all particles which hit the
scintillation detectors in each individual shower with observed the zenith and azimuth angles with help of the GEANT4 code.
In turn the code CORSICA-6.616 have been used to produce such particles in the atmosphere in these individual showers
and propagate them to detectors at the level of observation. Calculations have been carried out in terms of the QGSJET-2
and Gheisha-2002 models with the thinning parameter ${10}^{-8}$ for the primary protons and helium, oxygen and iron nuclei.
The energy of the most energetic extensive air shower observed at the Yakutsk array happened to be 200, 200, 180 and 165~EeV
with the values of the ${\chi}^2$ function per one degree of freedom 0.9, 1., 0.9 and 1.1 for the primary protons and helium,
oxygen and iron nuclei accordingly. Thus interpreting data in terms of the QGSJET-2 and Gheisha 2002 models we conclude that 
after the bump and decreasing
of the flux of the primary particles due to the Greisen, Zatsepin and Kuzmin effect that has been observed at the HiRes array
and at the Pierre Auger observatory there is a possible feature in the energy spectrum -- some increase in the flux at
energies 200 -- 300~EeV observed both at the Yakutsk array and at the AGASA array. Such possible feature may be understood
as the flux of heavy primary nuclei. It is also not excluded that some new component of the spectrum is observed or the
Lorentz invariance may be violated at such huge energies. As an alternative conclusion the models QGSJET-2 and Gheisha 2002
should be changed so as to produce much more muons in a shower.
\end{abstract}


\section{Introduction}

The energy of extensive air showers (EAS) observed at the Yakutsk array (YA) is usually estimated as follows.
First, for each individual shower the zenith and azimuth angles and coordinates of axis are determined.
Then the signal $s(600)$ is estimated from all scintillation detector readings. This signal is determined
by the energy deposited in a detector at a distance of 600~m from the shower core by all shower particles
(by electrons, positrons, gammas and muons). Then this signal is recalculated to the value it would have
in the vertical shower with the help of the average value of the attenuation length $<\lambda>$ which is estimated
by applying the constant integral intensity cut method \cite{1}. The calculated zenith-angle dependence is also
may be used but with average value of the attenuation length $<\lambda>$. But such procedure leads to large
uncertainties in energy estimate for the individual showers. First, in terms of model calculation it was
shown that the values of the attenuation length $<\lambda>$ estimated with the help of such procedure and calculated
for the average shower development differ considerably \cite{2,3}. Second, and it is a main source of uncertainty,
the individual showers may be generated by varies species of the primary particles. Besides, the real value of
the attenuation length $\lambda$ in individual vertical showers may differ from $\sim$ 200 up to $\sim$ 
2000~$g\cdot{cm}^{-2}$ \cite{4} due to
fluctuations in the longitudinal development. Indeed, the steepness of the individual cascade curve at the level
of observation (and the attenuation length $\lambda$) may vary very considerably due to fluctuations in the
points of first and subsequent interactions of the primary particle with atomic nuclei in the atmocphere.
Thus, the uncertainty of the signal $s(600)$ estimated for the vertical showers with the help of a factor such as
$\exp(-\Delta x/\lambda)$ may be very large, where $\Delta x$ is a slant depth. At last, the energy $E$ 
is estimated with help of the following formula

\begin{equation}
E=4.8\cdot{10}^{17}\cdot s(600),~eV.             \label{eq:1}
\end{equation}

\noindent
In the formula (1) uncertainties are missed for simplicity and it was suggested that the signal $s(600)$ is
proportional to the energy $E$ of a shower. This formula is based on the calibration of signals with help of the
Vavilov-Cherenkov radiation of a shower.
Again, this calibration has been carried out for the average value of a signal from some sample of showers.
For the individual vertical showers the numerical coefficient in (1) should vary due to fluctuations. Besides,
it should be mentioned that in terms of the model calculation the value of this numerical coefficient was
estimated as being 1.6 -- 1.7 times less than quoted in (1) \cite{5}.
Thus, any alternative methods of energy estimation are of interest. It was suggested
that readings of all detectors should be compared with calculated signals for a shower with the given values
of the zenith and azimuth angles \cite{6}. Calculations have been carried out for the giant shower observed at the
YA \cite{7} on the base of the original code. It was assumed in accordance with experimental data that this shower
consist mainly of muons and their deflections in the geomagnetic field have been taken into account. The energy
of this shower has been estimated as
$\sim 3\cdot {10}^{20}$~eV. In this paper calculations of signals for this giant shower at many points with 
different distances
from the shower core have been carried out for some sample of the individual showers induced by various primary
particles to take into account fluctuations in the longitudinal and lateral development. Then the
${\chi}^2$ method has been used to find out which of calculated individual showers agree best with data \cite{7}. As new
energy estimate are happened to be rather high some analyses of energy spectra observed at various arrays have been carried out
and the new interpretation of the energy spectrum has been suggested at ultra high energy region with the possible variable
contribution from the local sources.

\section{Method of simulations}

Simulations of the individual shower development in the atmosphere have been carried out with the help of code
CORSIKA-6.616 \cite{8} in terms of the models QGSJET2 \cite{9} and Gheisha 2002 \cite{10} with the weight 
parameter $\epsilon={10}^{-8}$ (thinning). The program GEANT4 \cite{11}
has been used to estimate signals in the scintillation detectors from electrons, positrons, gammas and muons.
The bank of detector responces has been calculated for electrons, positrons and gammas with energies in the 
interval 0.001--10~GeV
and muons with energies in the interval 0.3--1000~GeV which hit a detector at various the zenith angles 
(from $0^o$ up to ${60}^o$).
This bank of detector responces was used to estimate a signal in the scintillation detector when a shower particle 
hits it. The total area of $5\times 5$~${km}^2$ in the detector plane was divided into $201\times 201$ squares
with the side of 25~m. With the help of the code CORSIKA-6.616  the spread of shower particles in the detector plane 
has been estimated and the bank of detector responces has been used to calculate the signals in each square,
regarded as a detector. Thus, the matrix of $201\times 201$ detector
responces for each individual shower has been calculated. These matrixes of detector responces were 
calculated for individual showers with the same energy ${10}^{20}$~eV.
Calculations have been carried out for four species of the primary
particles (protons and nuclei of helium, oxygen and iron) with a statistics of four individual events for every
species of the primaries. Readings of the 31 scintillation
detectors have been used to search for the minimum of the function ${\chi}^2$ in the square with the width 
of 400~m and a center determined by data with a step of 1~m. 
These readings have been compared with calculated responces which were multiplied by the coefficient $C$. This 
coefficient changed from 0.1 up to 4.5 with a step of 0.1.
Thus, it was assumed, that the energy of a shower and signals in the scintillation detectors are proportional
to each other in some small interval.
New estimates of energy, coordinates of axis and values of the function 
${\chi}^2$ have been obtained for each individual shower.

The analysis of the energy spectra observed at various arrays has been carried out in the following way. 
The base universal spectrum $J_b(E)=A\cdot E^{-3.25}$ has been suggested mainly on assumption of data
\cite{12} at energies above ${10}^{17}$~eV with $A \approx 7.1 \cdot {10}^{28}$ $m^{-2} s^{-1} {sr}^{-1} 
{eV}^{2.25}$. All possible features of the energy spectrum of the primary
particles are considered relatively to this spectrum. Besides, the reference spectrum $J_r(E)$ has been
suggested as follows.
 For the energy we will use besides $E$ (in eV) 
additional notation $y=\lg (E/1~eV)$. In four energy intervals (i=1, 2, 3 and 4)
             $17.<y<18.65$, $18.65<y<19.75$, $19.75<y<20.01$ and $y>20.01$
the spectrum $J_r(E)$ has been approximated by the following exponent functions
         $$J_1(E)=A\cdot E^{-3.25},$$  
         $$J_2(E)=C\cdot E^{-2.81},$$ 
         $$J_3(E)=D\cdot E^{-5.1},$$ 
         $$J_4(E)=J_1(E)=A\cdot E^{-3.25}$$ accordingly.
Constants $C$ and $D$ may be expressed through $A$ and equations for $J_r(E)$ at the boundary points. For these 
four intervals we assume the reference spectrum as
                                          $$\lg z_i=\lg (J_i(E)/J_1(E)),$$
                                          where $i$=1, 2, 3, 4.
This reference spectrum is then represented as follows
         $$\lg z_1=0,$$ 
         $$\lg z_2=0.44\cdot (y-18.65),$$ 
         $$\lg z_3=0.484-1.85\cdot (y-19.75),$$
         $$\lg z_4=0$$
accordingly. We consider the spectrum $J_b(E)=A\cdot E^{-3.25}$ as universal up to highest energies. 
The first feature of the spectrum is suggested to be considered as some excess at energies $18.65\le y \le 20.01$. The left side
of this exess is approximated as $J_2(E)-J_1(E)$ while the right side as $J_3(E)-J_1(E)$. The possible
second feature at energies (2--3)$\cdot {10}^{20}$~eV will also be discussed.
Results of the spectra $J(E)$ observed at various arrays have been expressed as
                                                $$\lg z=\lg (J(E)/J_1(E))$$
and are shown in comparison with the reference spectrum.

 \begin{figure}[!t]
  \centering
  \includegraphics[width=8cm]{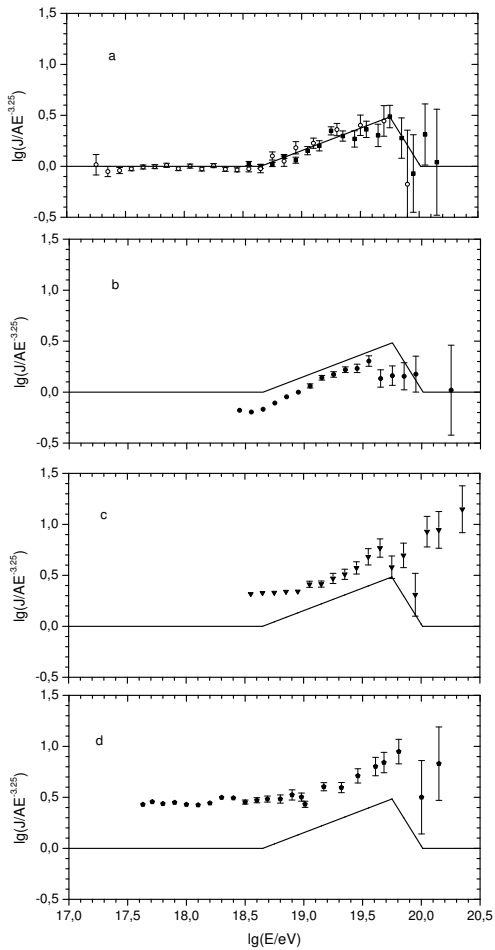}
  \caption{The energy spectra observed at various arrays and the reference spectrum (solid line): 
  a) -- \cite {12}, (b) -- \cite {16}, (c) -- \cite {13},
  (d) -- \cite {17}}
  \label{simp_fig}
 \end{figure}

\section{Results of energy estimations and analysis of the energy spectrum}

The 16 various values of energy
estimates for 16 individual simulated showers with different  values of the function ${\chi}^2$ have been 
obtained for the same sample of the 31 experimental readings of the observed giant shower \cite {7}. As
the giant shower is very inclined muons make main contribution to signals in the scintillation detectors. 
The energy estimates are minimal and change inside
the interval (1.6--1.75)$\cdot {10}^{20}$~eV with the value $\sim 1.1$ of the ${\chi}^2$ function per one degree 
of freedom for the
iron nuclei primaries which produce more muons than the proton primaries . For the proton
and helium nuclei primaries energy estimates are maximal and change inside the interval 
(1.8--2.4)$\cdot {10}^{20}$~eV with the value $\sim 0.9$ of the ${\chi}^2$ function per one degree of freedom. 
For the oxygen nuclei primaries the energy estimates are in the interval (1.8--2)$\cdot {10}^{20}$~eV which is
between intervals for proton and iron nuclei primaries. It should be mentioned that only muons contribute
$\sim 80\%$ of the total signal and therefore
energy estimate increases up to $2.8\cdot {10}^{20}$ eV for the proton primaries if only muons are taken into 
account as in \cite{7} disregarding the contribution of electrons, positrons and gammas.
New coordinates of shower axis vary from the experimental one by some dozen of meters. So we can make main conclusion
that in terms of the QGSJET2 \cite {9} and Gheisha 2002 \cite {10} models and the proton primaries there are extensive 
air showers with energies $\sim 2\cdot {10}^{20}$~eV.
This estimate may be decreased up to $\sim 1\cdot {10}^{20}$~eV only in terms of some new model which produces twice 
as much muons as \cite {9} and \cite {10}.
It is not easy because muons are produced in low energy region where the physics of interactions is known. But the energy 
estimate decreases up to $\sim 1.6 \cdot {10}^{20}$~eV for the iron nuclei as the primary particles. It should be
remind that such giant showers have been also observed in \cite {13}. It looks as a some contradiction to the 
suggestion by
Greisen, Zatsepin and Kuzmin (GZK) \cite {14,15}. Just on the contrary, the observations \cite {12,16} show no giant showers in 
accordance with
the GZK prediction but not with data \cite {13,17}. It is evidently that all world data should be understood. 
Of course, some uncertainties in energy estimates may exist. 

But it is also worth-while, to consider some new idea about the energy spectrum at
ultra high energies. Usually, this spectrum is considered as universal and stationary \cite {18}. It is possible if many
uniformly distributed sources contribute to the spectrum. In case of the near-by sources \cite {19} distributed 
anisotropic-ally
their contribution to the different intervals of the spectrum may be variable. If a power of local sources is not high one
source may give a contribution to the number of observed events only once in many years. So, various sources may contribute
to the spectrum at different time of an exposure. Due to deflection in magnetic
fields arrival directions of showers differ from directions to local sources. If the number of sources is not high their
contribution to the spectrum would be very variable and even chaotic. The primary particles from such local sources may
be considered as some variable excess above the suggested universal spectrum $J_b(E)$. Decreasing of the flux of the primary
particles due to the GZK effect should probably be considered relatively to this universal spectrum but not relatively the
observed bump as in \cite {12,16}. Evidently,
it is a difficult problem. 
Of course, the flux of particles at exess energies is also supressed by the GZK effect.
Fig. 1 illustrates our suggestion. Data $\lg z=\lg (J(E)/J_1(E))$ observed at various arrays are
shown in Fig. 1 as follows: (a) -- \cite {12} (open circles -- HiRes2, solid circles -- HiRes1), (b) -- \cite {16} 
(solid circles), (c) -- \cite {13} (solid triangles) and (d) -- \cite {17} (solid pentagons). 
The reference spectrum is also shown on all Figures. As
it was expected the data \cite {12} agree very well with the reference spectrum. The data \cite{16} are below this reference
spectrum, and
the data \cite {13} and \cite {17} are much above it. What possible features are seen in Fig. 1? First, no predicted dip \cite{18} 
at energy ${10}^{18}$~eV
is seen. Just the contrary is true: in this energy region the exponents of all spectra are approximately the same with
a good accuracy.
Second, all data show some excess in the energy interval (5--10)$\cdot {10}^{19}$~eV (so called "bump" \cite{18}) which 
may be considered as a
contribution from the local sources. The observation of the anisotropy of arrival directions of showers \cite{16,19} supports this
statement. At last, no dramatic fall of the flux of the primary particles relatively to the reference spectrum is seen at
energies above ${10}^{20}$~eV. Indeed, due to \cite{12} the number of expected events with energies above 
$6.3\cdot {10}^{19}$~eV is equal to
43.2 but only 13 events were observed. According to the reference spectrum the number of expected events is calculated as
16 that with the Poisson fluctuation taken into account agrees with the observed number. Due to \cite{16} numbers of 
expected
events with energies above $6\cdot {10}^{19}$~eV and ${10}^{20}$~eV are equal to 167 and 35 accordingly while only 
69 and 1 events were
observed. The reference spectrum gives 137 and 7 events accordingly. The disagreement decreased but not vanished. This fall
may be regarded as the observation of the GZK supression of the flux of the primary particles. It may be
commented that data \cite{16} include two empty bins which should contain 6 events. The Poisson probability that no one was
observed equals to $\sim 2.5\cdot {10}^{-3}$. It should also be mentioned that the intensity of the reference spectrum is 
$\sim 1.5$ times higher
than used in \cite{16}. Besides, some uncertainties in energy estimates are possible as we believe due to the constant integral
intensity cut method \cite{1} which disregarded fluctuations. The data \cite{13} illustrate possibly a contribution of local
variable sources at energies above ${10}^{20}$~eV while some uncertainties in energy estimates or aperture may be of
importance at energies below $5\cdot {10}^{19}$~eV. The same comments may be addressed to data \cite{17} with additional 
remark that calculated energy estimate is 1.6 times less than used in data. Besides, data \cite{13,17} show probably the
second variable excess at energies (2--3)$\cdot {10}^{20}$~eV. This second variable exess may be regarded as a contribution of heavy nuclei
to the flux of the primary particles from the local sources. It is also not excluded that the Lorentz invariance may be 
violated at such huge energies \cite{20}, \cite{21}, \cite{22}. The intensity of local sources on the Earth may be estimated
 as integral on differences $J_2-J_1$ and $J_3-J_1$ accordingly in case of the first exess. 
This integral is estimated as $I \approx 4\cdot {10}^{-14}$~$m^{-2} s^{-1} {sr}^{-1}$. If we assume a distance
to local sources as $R \sim 30$~Mpc, a typical energy $E \sim {10}^{19}$~eV and an angle of emission as $\sim 1$~sr then
we obtain the power of all local sources as $3\cdot {10}^{33}$~W which is in agreement with estimates in \cite{23,24}.

\section{Conclusion}

The new method has been suggested to estimate energy of extensive air showers by comparison all detector
readings with calculated signals for a sample of individual events induced by various primary particles. 
Simulations of the individual shower development in the atmosphere have been carried out with the help of code
CORSIKA-6.616 \cite{8} in terms of the models QGSJET2 \cite{9} and Gheisha 2002 \cite{10} with the weight 
parameter $\epsilon={10}^{-8}$ (thinning). The program GEANT4 \cite{11}
has been used to estimate signals in the scintillation detectors from electrons, positrons, gammas and muons.
New estimates of energy of the giant air shower
observed at YA \cite{7} have been calculated in terms of the QGSJET2 \cite{9} and Gheisha 2002 \cite{10} 
models as $E\sim 2\cdot {10}^{20}$~eV for
the proton primaries and $E\sim 1.7\cdot {10}^{20}$~eV for the primary iron nuclei. 
The base universal spectrum such 
as $J_b=A\cdot E^{-3.25}$
have been suggested at energies above ${10}^{17}$~eV. It was also suggested that some possible
local sources may produce variable contribution to the different regions of the energy spectrum at super high energies.
The fall of the flux of the primary particles due to the GZK effect should be considered relatively the
some base spectrum. Possibly, the second exess at energies (2--3)$\cdot {10}^{20}$~eV has been observed in \cite{13,17}.

\section{Acknowledgements}

Moscow authors thank RFBR (grant 07-02-01212) and G.T. Zatsepin LSS
 (grant 959.2008.2) for support.\\
 Autors from Yakutsk thank RFBR (grant 08-02-00348) for support.


\begin{thebibliography}{24}
 \bibitem{1} G.~Clark et al., Proc. 8th ICRC. Jaipur. {\bf 4}, 65 (1963).
 \bibitem{2} L.G.~Dedenko., Proc. 14th ICRC. Munchen. {\bf 8}, 2857 (1975).
 \bibitem{3} L.G.~Dedenko et al.,  Nucl. Phys. B (Proc. Suppl.) {\bf 136}, 12 (2004).
 \bibitem{4} L.G.~Dedenko et al.,  Bull. of Russian Acad. of Sci. Phys., {\bf 66}, 1603, (2002).
 \bibitem{5} L.G.~Dedenko et al.,  Yad. Fiz., {\bf 70}, 1806 (2007).
 \bibitem{6} E.E.~Antonov et al.,  JETP Lett. {\bf 69}, 614 (1999).
\bibitem{7}  N.N.~Efimov et al., Proc. Int. Workshop on Astrophysical  Aspects of the Most Energetic Cosmic Rays, Kofu, Japan. 20 (1990).
\bibitem{8}  D.~Heck et al., Forschungszentrum Karlsruhe Thechnical Report No. 6019, 1998.
\bibitem{9}  S.S.~Ostapchenko et al., Nucl. Phys. B (Proc. Suppl.) {\bf 151}, 143 (2006).
\bibitem{10} H.~Fesefeldt, Report PITHA-85/02, RWTA, Aachen (1985).
\bibitem{11} The GEANT4 Collab., http:/www.info.cern.ch/asd/geant4.html.
\bibitem{12} R.U.~Abbasi et al. ,(High Resolution Fly's Eye Collaboration). PRL.{\bf 100}, 101101 (2008).
\bibitem{13} M.~Takeda et al., Astropart. Phys. {\bf 19}, 447 (2003).
\bibitem{14} K.~Greisen. Phys. Rev. Lett. {\bf 16}, 748 (1966).
 \bibitem{15} G.T.~Zatsepin and V.A.~Kuzmin,  JETP Lett. {\bf 4}, 78 (1966).
  \bibitem{16} J.~Abraham et al.,(The Pierre Auger Collaboration). PRL.{\bf 101}, 061101 (2008).
 \bibitem{17} V.P.~Egorova et al.,  Nucl. Phys. B (Proc. Suppl.) {\bf 136}, 3 (2004).
\bibitem{18} V.~Berezinsky et al., Phys. Rev. D {\bf 74}, 043005 (2006).
\bibitem{19} J.~Abraham et al.,(The Pierre Auger Collaboration). Science {\bf 318}, 939 (2007).
\bibitem{20} S.~Coleman and S.L.~Glashow, Phys. Rev. D {\bf 59}, 116008 (1999).
\bibitem{21} E.E.~Antonov et al.,  JETP Lett. {\bf 73}, 506 (2001).
\bibitem{22} S.T.~Scully, F.W.~Stecker, Astropart. Phys. {\bf 31}, 220 (2009).
\bibitem{23} L.G.~Dedenko et al., arXiv:astro-ph/0703015v1 [astro-ph] (1 March 2007).
\bibitem{24} L.G.~Dedenko et al., arXiv:astro-ph/0811.0722v1 [astro-ph] (5 Nov 2008).
\end{thebibliography}
\end{document}